\documentclass{aip-cp}

\usepackage[numbers]{natbib}
\usepackage{rotating}
\usepackage{graphicx}

\begin{document}

\title{Note on neutron star equation of state in the light of GW170817}
\author[aff1]{Ang Li\corref{cor1}}
\author[aff1,aff2]{Zhen-Yu Zhu}
\eaddress{zhenyuz@stu.xmu.edu.cn}
\affil[aff1]{Department of Astronomy, Xiamen University, Xiamen 361005, China}
\affil[aff2]{Institute for Theoretical Physics, Frankfurt am Main 60438, Germany}
\corresp[cor1]{liang@xmu.edu.cn}

\maketitle

\begin{abstract}
From the very first multimessenger event of GW170817, clean robust constraints can be obtained for the tidal deformabilities of the two stars involved in the merger, which provides us unique opportunity to study the equation of states (EOSs) of dense stellar matter. In this contribution, we employ a model from the quark level, describing consistently a nucleon and many-body nucleonic system from a quark potential. We check that our sets of EOSs are consistent with available experimental and observational constraints at both sub-nuclear saturation densities and higher densities. The agreements with ab-initio calculations are also good. Especially, we tune the density dependence of the symmetry energy (characterized by its slope at nuclear saturation $L$) and study its influence on the tidal deformability. The so-called $QMF18$ EOS is named after the case of $L=40~\rm MeV$, and it gives $M_{\rm TOV} =2.08~M_\odot$ and $R= 11.77~\rm km$, $\Lambda=331$ for a $1.4\,M_\odot$ star. The tidal signals are demonstrated to be insensitive to the uncertainty on the crust-core matching, despite the good correlation between the symmetry energy slope and the radius of the star.
\end{abstract}

\section{INTRODUCTION}

Thanks to the development of many-body theories of nuclear matter, there is a possibility that the long-standing, open problem of the equation of state (EOS) of dense matter can be understood, by confronting laboratory measurements of nuclear properties \& reactions (e.g., RIBLL at HIRFL, HIRA at NSCL, CEE at HIAF, CBM at FAIR) and observations in astronomy (e.g., HXMT~\cite{hxmt}, eXTP~\cite{extp1,extp2}, SVOM~\cite{svom}, SKA~\cite{ska}, FAST~\cite{fast}, Urumqi, Lijiang~\cite{lijiang}, NICER~\cite{nicer}), especially after the recent multimessenger observations of neutron-star (NS) merger GW170817~\cite{2017PhRvL.119p1101A,2018PhRvL.121p1101A}. 

The tidal deformability $\Lambda$ describes the amount of induced mass quadrupole moment when reacting to a certain external tidal field \citep{1992PhRvD..45.1017D,2009PhRvD..80h4035D}. It is normalized with a factor of $R^5$ from the second Love number $k_2$, being $R$ the NS radius. $k_2$ also has a dependence on $R$ (see, e.g., \cite{2010PhRvD..81l3016H,2010PhRvD..82b4016P, 2018ApJ...862...98Z, 2018PhRvD..97h3015Z}). Following the tidal deformability observation of the GW170817 event at the premerger stage, robust lower limits can be put on the radius of merging stars' radii, which are around 10.7 km~\cite{2017ApJ...850L..34B}. A more detailed result from LIGO-Virgo Collaboration is $11.9^{+1.4}_{-1.4}$ km at the $90\%$ level~\cite{2018PhRvL.121p1101A}. 

On the other hand, it has been established that the star radius is rather sensitive to the symmetry energy (essentially its slope $L$) with the maximum mass only slightly modified~(see, e.g.,~\cite{2001ApJ...550..426L,2004Sci...304..536L,2006PhLB..642..436L}). A smaller $L$ (softer symmetry energy) leads to a smaller radius. Therefore, it is meaningful to study the relation of the tidal deformability with the uncertain $L$ parameter, i.e., the symmetry energy slope at nuclear matter saturation. We will examine in details in this contribution how the presently uncertain symmetry energy slope influences the tidal deformability of GW170817-like events. We use the matter state model from the quark level and suppose the merging stars are both NSs without strangeness phase transitions. The employed model has the advantage to tune consistently only the slope of interest with the other saturation properties fixed at their empirical values, respectively. Brief discussions are also made on the effect of crust-core matching and the possibility of drawing information on (inner-)crust EOS from GW signals.

\section{MODEL}

\begin{figure*}[h]
\begin{minipage}{18pc}
\includegraphics[width=18pc]{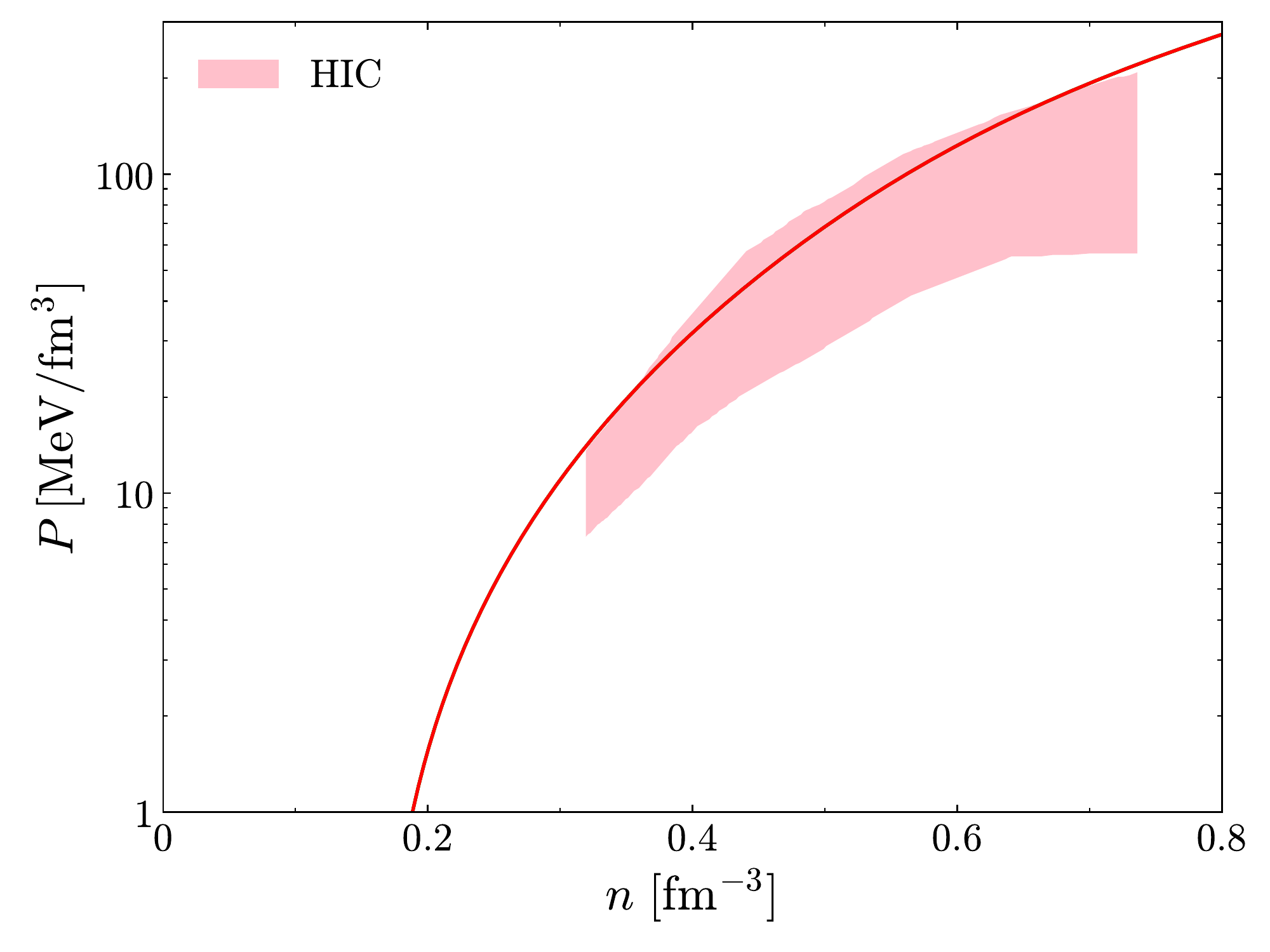}
\caption{Pressure as a function of nucleon number density for SNM, together with the constraint from collective flow in heavy-ion collisions~\cite{2002Sci...298.1592D}. Note that the results for different values of the symmetry energy slope $L$ overlap in this plot. The SNM EOS is compatible with the flow constraint. Adapted from \cite{2018ApJ...862...98Z}.}
\end{minipage}
\hspace{1pc}%
\begin{minipage}{18pc}
\includegraphics[width=18pc]{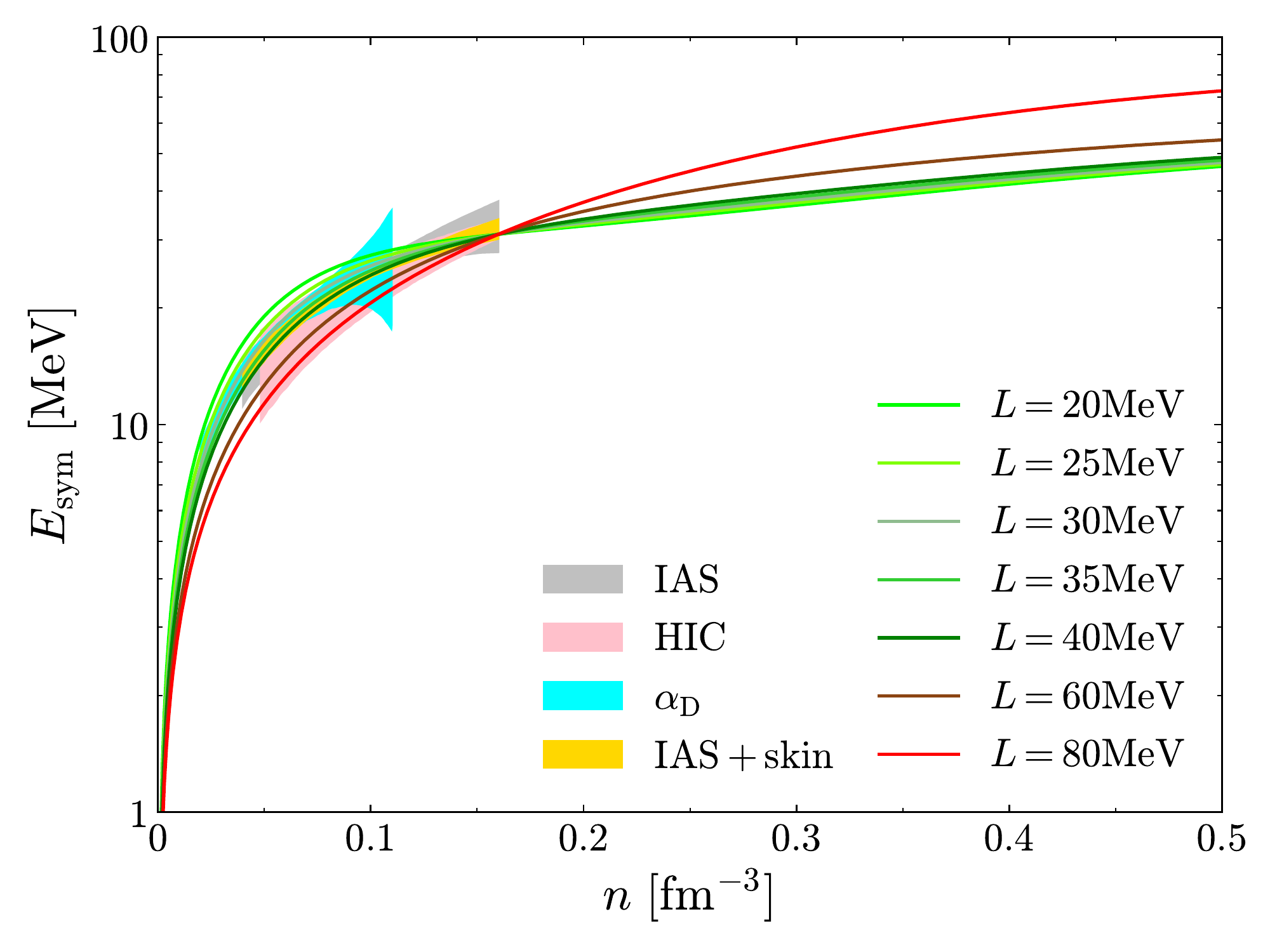}
\caption{Symmetry energy as a function of nucleon number density with different values of symmetry energy slope $L$. Colorful shadow regions represent the constraints from isobaric analog states (IAS) and from transport in heavy-ion collisions~\cite{2009PhRvL.102l2701T}, electric dipole polarizability in $^{\rm 208}$Pb ($\alpha_D$)~\cite{2015PhRvC..92c1301Z}, IAS and neutron skins (IAS+skin)~\cite{2014NuPhA.922....1D}, respectively. For different $L$, the symmetry energy results are all consistent with various nuclear experiments, with the cases around $L = 40$ MeV agreeing better. Adapted from \cite{2018ApJ...862...98Z}.}
\end{minipage}
\end{figure*}

The theoretical model we employ here is the quark mean-field (QMF) model (see e.g.,~\cite{2018ApJ...862...98Z,2000PhRvC..61d5205S,2002NuPhA.707..469S,2014PhRvC..89b5802H,2014PTEP.2014a3D02H,2016PhRvC..94d4308X,2018PhRvC..97c5805Z} and its recent extension \cite{2018arXiv180504678Z}). The model starts with a flavor independent two-parameter potential, $U(r)=\frac{1}{2}(1+\gamma^0)(ar^2+V_0)$, confining the constituent quarks inside a nucleon. The Dirac equation of the confined quarks is written as
\begin{eqnarray}
[\gamma^{0}(\epsilon_{q}-g_{\omega q}\omega-\tau_{3q}g_{\rho q}\rho)-\vec{\gamma}\cdot\vec{p} -(m_{q}-g_{\sigma q}\sigma)-U(r)]\psi_{q}(\vec{r})=0,
\end{eqnarray}
where $\psi_{q}(\vec{r})$ is the quark field, $\sigma$, $\omega$, and $\rho$ are the classical meson fields. $g_{\sigma q}$, $g_{\omega q}$, and $g_{\rho q}$ are the coupling constants of $\sigma, ~\omega$, and $\rho$ mesons with quarks, respectively. $\tau_{3q}$ is the third component of isospin matrix. This equation can be solved exactly and its ground state solution for energy is
\begin{eqnarray}
(\mathop{\epsilon'_q-m'_q})\sqrt{\frac{\lambda_q}{a}}=3,
\end{eqnarray}
where $\lambda_q=\epsilon_q^\ast+m_q^\ast,\ \mathop{\epsilon'_q}=\epsilon_q^\ast-V_0/2,\ \mathop{m'_q}=m_q^\ast+V_0/2$. The effective single quark energy is given by $\epsilon_q^*=\epsilon_{q}-g_{q\omega}\omega-\tau_{3q}g_{q\rho}\rho$ and the effective quark mass by $m_q^\ast = m_q-g_{\sigma q}\sigma$ with the quark mass $m_q$ = 300 MeV. The zeroth-order energy of the nucleon core $E_N^0=\sum_q\epsilon_q^\ast$ can be obtained by solving Eq.~(1). Corrections due to center-of-mass motion $\epsilon_{c.m.}$, quark-pion coupling $\delta M_N^\pi$, and one gluon exchange $(\Delta E_N)_g$ are included to obtain the nucleon mass, see details in \cite{2018ApJ...862...98Z}. With these corrections on the energy, we can then determine the mass of nucleon in medium:
\begin{eqnarray}
M^\ast_N=E^{0}_N-\epsilon_{c.m.}+\delta M_N^\pi+(\Delta E_N)_g.
\end{eqnarray}
The nucleon radius is written as
\begin{eqnarray}
\langle r_N^2\rangle = \frac{\mathop{11\epsilon'_q + m'_q}}{\mathop{(3\epsilon'_q + m'_q)(\epsilon'^2_q-m'^2_q)}}.
\end{eqnarray}
The potential parameters ($a$ and $V_0$) are determined from reproducing the nucleon mass and radius in free space, namely $M_N = 939$ MeV and $r_N = 0.87$ fm. 

Then nuclear matter is described by point-like nucleons and mesons interacting through exchange of $\sigma,~\omega,~\rho$ mesons. The cross coupling from $\omega$ meson and $\rho$ meson is also included. The calculation of confined quarks gives the relation of effective nucleon mass as a function of $\sigma$ field, which defines the $\sigma$ coupling with nucleons (depending on the parameter $g_{\sigma q}$). The meson coupling constants are fitted by reproducing the empirical saturation properties of nuclear matter.  $m_{\sigma} = 510~\rm{MeV}$,~$m_{\omega}=783~\rm{MeV}$, and $m_{\rho}=770~\rm{MeV}$ are the meson masses. The QMF framework describes consistently a nucleon and many-body nucleonic system from a quark potential. 

The Lagrangian for describing nuclear matter is written as:
\begin{eqnarray}
\mathcal{L}& = & \overline{\psi}\left(i\gamma_\mu \partial^\mu - M_N^\ast - g_{\omega N}\omega\gamma^0 - g_{\rho N}\rho\tau_{3}\gamma^0\right)\psi  -\frac{1}{2}(\nabla\sigma)^2 - \frac{1}{2}m_\sigma^2 \sigma^2 - \frac{1}{3} g_2\sigma^3 - \frac{1}{4}g_3\sigma^4 \nonumber \\ 
& & + \frac{1}{2}(\nabla\rho)^2 + \frac{1}{2}m_\rho^2\rho^2 + \frac{1}{2}(\nabla\omega)^2 + \frac{1}{2}m_\omega^2\omega^2 + \frac{1}{2}g_{\rho N}^2\rho^2 \Lambda_v g_{\omega N}^2\omega^2, 
\end{eqnarray}
where $g_{\omega N}$ and $g_{\rho N}$ are the nucleon coupling constants for $\omega$ and $\rho$ mesons. From the quark counting rule, we obtain $g_{\omega N}=3g_{\omega q}$ and $g_{\rho N}=g_{\rho q}$. 
There are six parameters ($g_{\sigma q}, g_{\omega q}, g_{\rho q}, g_2, g_3, \Lambda_v$) in the Lagrangian. Recently, several new QMF parameter sets have been fitted by reproducing the saturation density $n_0=0.16~\rm fm^{-3}$ and the corresponding values at saturation point for the binding energy $E/A=-16~\rm MeV$, the incompressibility $K=240~\rm MeV$~\cite{2006EPJA...30...23S,2010JPhG...37f4038P}, the symmetry energy $E_{\rm sym}=~31~\rm MeV$~\cite{2013PhLB..727..276L}, the symmetry energy slope $L= 20-80~\rm MeV$~\cite{2013PhLB..727..276L,2009PhRvL.102l2502C} and the effective mass $M_N^\ast=0.77$. We refer to \cite{2018ApJ...862...98Z} for the detailed values of model parameters. We report in Figure 1 and Figure 2 the resulting EOS in symmetric nuclear matter (SNM) and the symmetry energy, respectively. We see overall good agreements of them with various laboratory nuclear experiments. The slope $L$ changes evidently the density dependence of the symmetry energy, and different outcomes deviate more at higher densities. The current laboratory constraints seem favor the cases around $L = 40$ MeV.  

\begin{figure}[h]
  \centerline{\includegraphics[width=300pt]{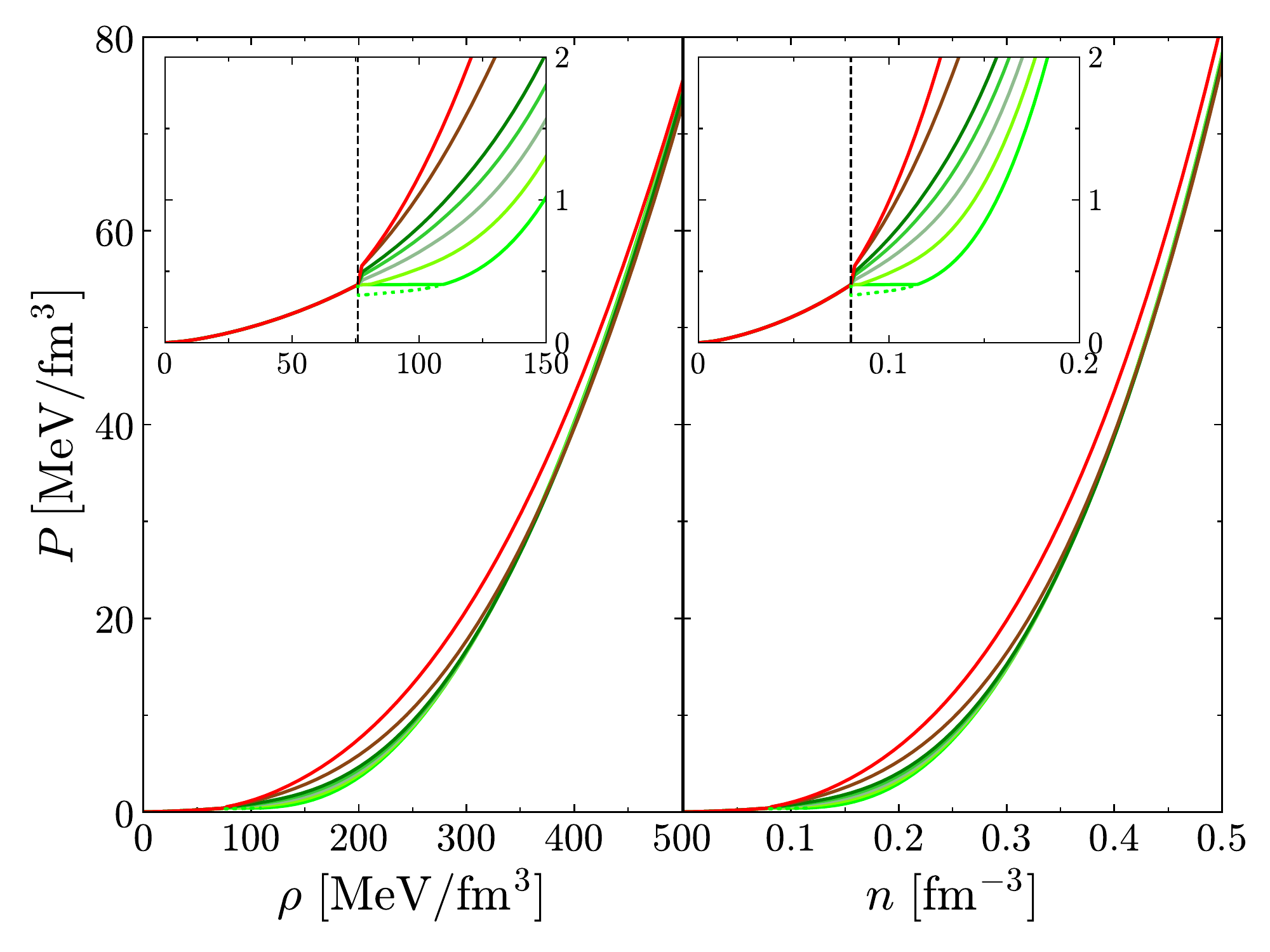}}
  \caption{Pressure of neutron star matter as a function of both energy density and number density, with different values of symmetry energy slope $L$. The colour coding is the same with Fig 2. The insets show the crust-core matching details. The core EOS is from the present QMF calculation. The inner (outer) crust EOS is the usual NV (BPS) one~\cite{bps,nv}. We keep the energy density as increasing functions of $P$, using simple horizontal cutoff if necessary, for example in the cases of $L= 20,~25~\rm MeV$.}
\end{figure}

\section{RESULTS AND DISCUSSIONS}

For calculating the EOS of NS matter, we introduce beta-equilibrium and charge neutrality condition between nucleons and leptons. Figure 3 presents the resulting EOS of NS matter with different values of symmetry energy slope $L$. We glue different core EOSs to the same NV + BPS crust EOS~\cite{bps,nv}, keeping the pressure $P$ as increasing functions of the energy density $\rho$. The matching procedure can be different as discussed in \cite{2016PhRvC..94c5804F}, and the proper way to do is using Maxwell construction, guaranteeing that the pressure is an increasing function of both the density and the chemical potential. We plan to further improve this part after finishing developing the unified EOS in our model. Presently it is important to note that there may be no thermodynamic consistency, or the speed of sound may be unphysical in some $L$ cases. Since it is the $P(\rho)$ function enters the TOV equations, we make use of the present collections of the star EOSs, and move forwards to discuss the resulting mass-radius relations, as well as the tidal deformability of merging system, focusing on the effects of both the slope parameter and the behaviour of the matching interface.

\begin{figure}[h]
  \centerline{\includegraphics[width=320pt]{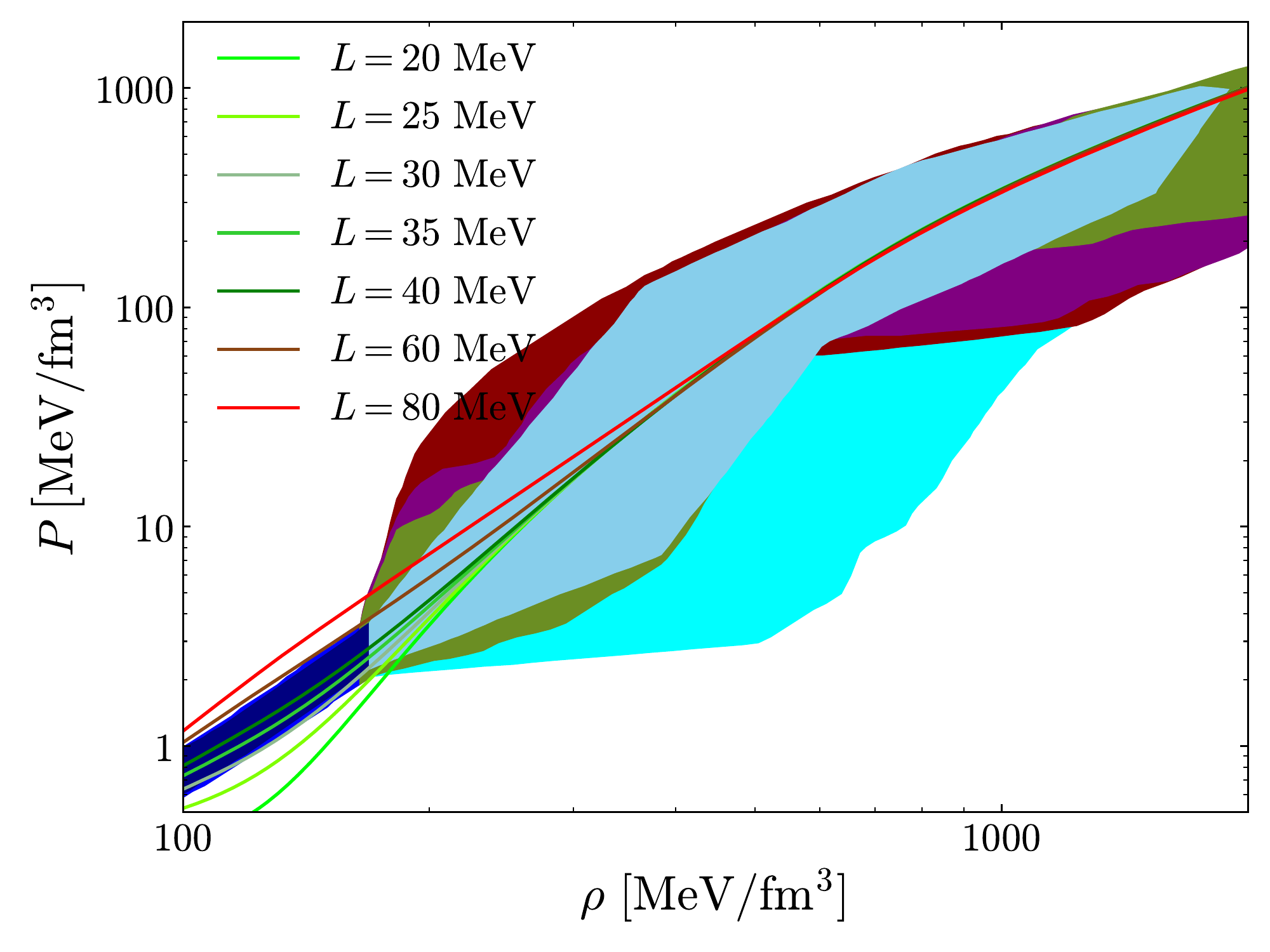}}
  \caption{NS EOSs within QMF with different values of symmetry energy slope $L$, to be compared with the favored regions from ab-initio calculations at subsaturation density in chiral effective field theory and at very high density in perturbative QCD. The blue region at lower densities is from the calculation of pure neutron matter incorporated with beta equilibrium. The lighter blue region is the envelope of its general polytropic extensions that are causal and support a neutron star of $\sim 2 M_{\odot}$~\cite{2010Natur.467.1081D,2013Sci...340..448A,2016ApJ...832..167F}. They are both from Hebeler et al. 2013~\cite{2013ApJ...773...11H}. Other regions are from Annala et al. 2018~\cite{2018PhRvL.120q2703A}, with color coding same with \cite{2018PhRvL.120q2703A} and will be explained later in Figure 5. 
}
\end{figure}    
    
Before doing so, we further compare our star EOSs with first-principle calculations of nuclear EOS in chiral effective field theory and in perturbative QCD. Various preferred regions are included in our Figure 4, from Hebeler et al. 2013~\cite{2013ApJ...773...11H} and Annala et al. 2018~\cite{2018PhRvL.120q2703A}. We see immediately that the very small and very large cases of $L= 20, 25, 80 \rm MeV$ are not compatible with the lower density band based on chiral effective field theory~\cite{2013ApJ...773...11H}. The allowed $L$ values in our QMF model may be in the range of $\sim 30-60~\rm MeV$ from this neutron matter constraint. And the cases of $L=30-60~\rm MeV$ also locate within the uncertainty bands of its general polytropic extensions over the entire density range. The slope parameter affects more evidently the low-density ranges than the high-density ranges. The four cases of $L=30-60~\rm MeV$ lie within the green $\Lambda < 400$ region. Those can be more clearly seen in Figure 5 in the plots of the NS mass-radius relations. It is expected that the TOV mass of the star hardly changes with changing $L$, but there is a strong positive correlation between the slope parameter and the radius of a $1.4\,M_\odot$ star (see more discussion in e.g., \cite{2018PhRvC..98f5804H,2018PhRvL.121f2701L} and a small dependence is found, however, in \cite{2018PhRvC..98f5804H}). $L$ parameter does not affect much $\Lambda$, if we consider the preferred cases of $L=30-60~\rm MeV$.

\begin{figure}[h]
  \centerline{\includegraphics[width=320pt]{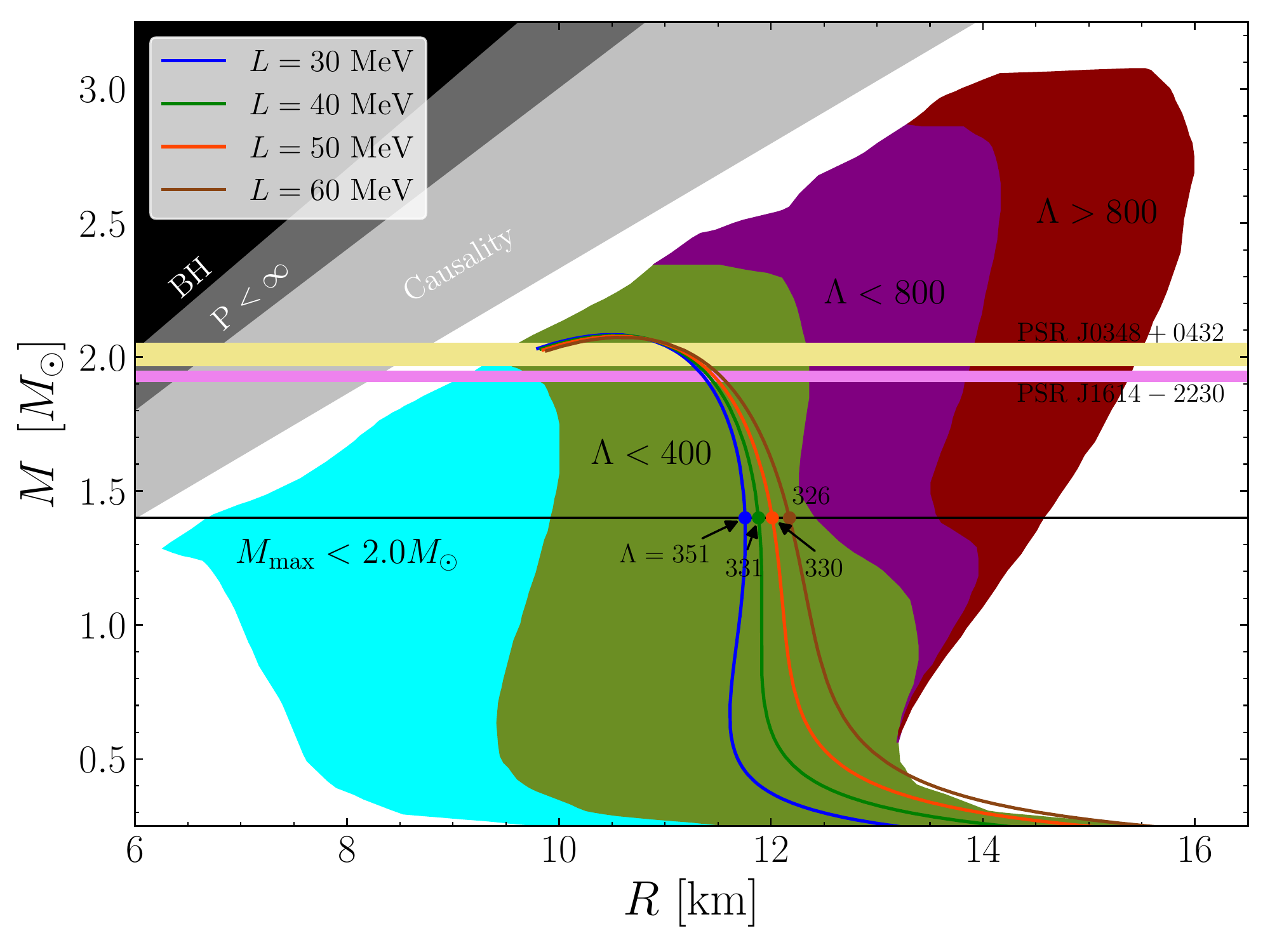}}
  \caption{Mass-radius relation of neutron star with different values of symmetry energy slope $L$, to be compared with the results from Annala et al. 2018~\cite{2018PhRvL.120q2703A}. $M_{\rm max}$ stands for the maximum gravitational mass in static, or the TOV mass. $\Lambda$ is the dimensionless tidal deformability for a $1.4\,M_\odot$ star. The mass measurements of two heavy pulsars are also shown~\citep{2013Sci...340..448A,2010Natur.467.1081D,2016ApJ...832..167F}, as well as the specific values of $\Lambda$ for four $L$ cases. The shaded regions show the black hole limit, the Buchdahl limit and the causality limit, respectively. Adapted from \cite{2018ApJ...862...98Z}.}
\end{figure}

\begin{figure*}[h]
\includegraphics[width=21pc]{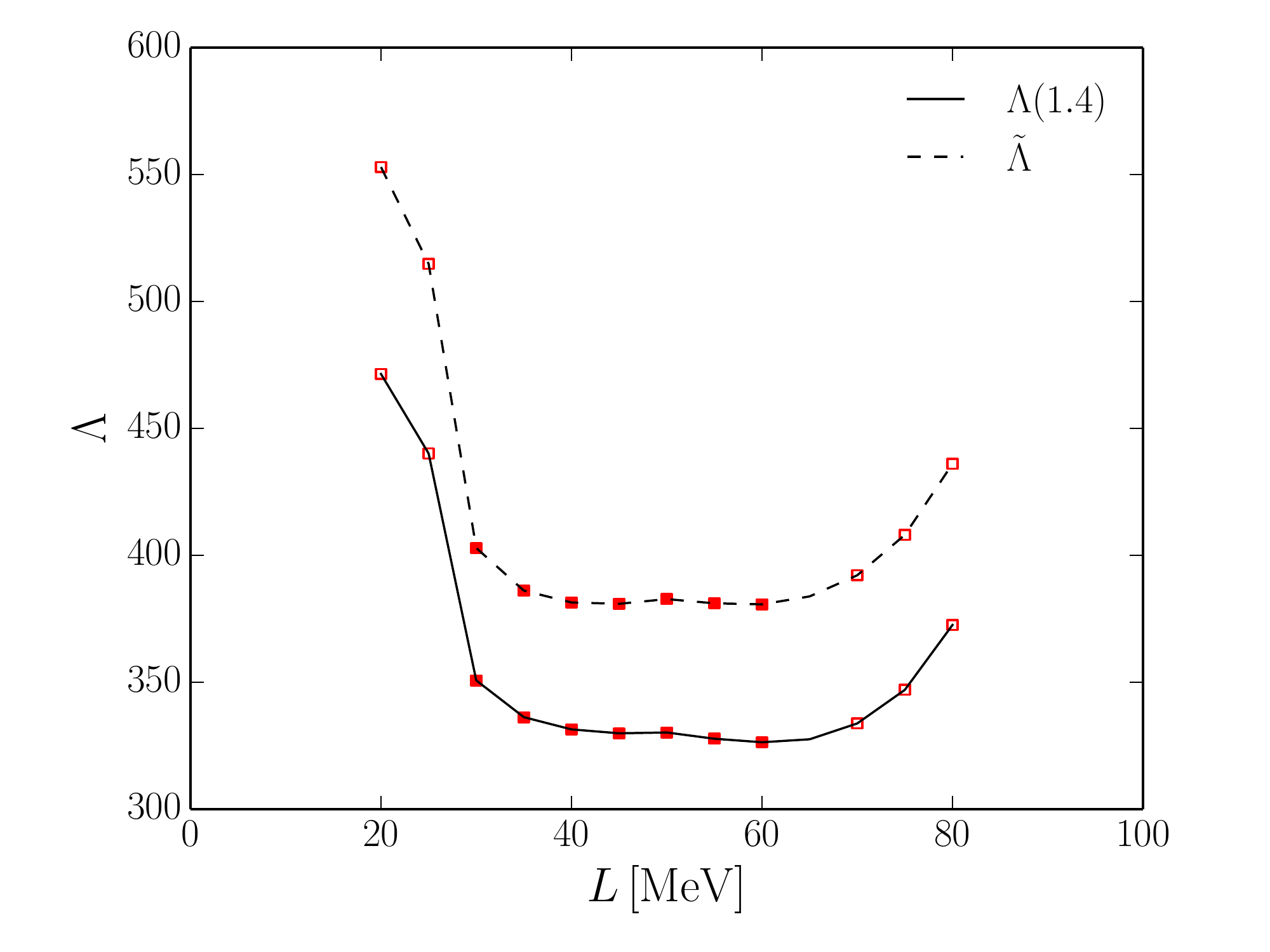}\hspace{0pc}%
\begin{minipage}[b]{17pc}
\caption{$L$ dependence of tidal deformability: (In solid) the tidal deformability for a $1.4\,M_\odot$ star and (in dashed) the mass-weighed tidal deformability $\tilde{\Lambda}$ of a binary system with a chirp mass of 1.188$\,M_{\odot}$, and mass ratio of 0.7~\citep{2017PhRvL.119p1101A}. Taken from \cite{2018ApJ...862...98Z}.}\end{minipage}
\end{figure*}

To understand better the relation between $L$ and $\Lambda$, we present in Figure 5, for more $L$ values, the results of both $\Lambda$ and the measured mass-weighed tidal deformability ($\tilde{\Lambda}$). A chirp mass of 1.188$\,M_\odot$ and mass ratio of 0.7 ~\cite{2017PhRvL.119p1101A} are employed for the calculation of $\tilde{\Lambda}$ in a binary system. Since the mass-weighed tidal deformability is expected to be very weakly dependent on the mass ratio (see e.g., \cite{2018ApJ...852L..29R}), considering one mass ratio case should be representative enough for analysis. In the more reliable ranges of $L=30-60~\rm MeV$ within QMF, neither $\Lambda(1.4)$ nor $\tilde{\Lambda}$ shows good correlation with $L$. Similar results are also found by Lim \& Holt 2018~\cite{2018PhRvL.121f2701L}. Previously, we discussed this unexpected point in \cite{2018ApJ...862...98Z} mainly using the relevance of a crust of a small star with a mass $1.4\,M_\odot$ and the possibility of drawing information on the (inner-)crust EOS from GW signals. However, the analysis of Gamba et al.~\cite{2019arXiv190204616G} argued negative, where they sampled the symmetry energy (its slope) in the interval of $30-40~\rm MeV$ ($30-70~\rm MeV$), reanalyzed the data of GW170817 by adding the level of uncertainty coming from the choice of the crust structure model. In their study, it is also true that the choice of the crust EOS affects the radii of the NSs in the coalescence ($\sim 3\%$, and about 0.3 km), but the tidal parameters are not sensitive to the EOS at low crust densities. The GW measurements mainly probe the high density EOS in NSs' cores. The low sensitivity of $L$ parameter to the tidal signal can then be understood, considering it only affects the lower density EOS as seen in Fig 4.

We have proposed a new $QMF18$ EOS in our previous study~\cite{2018ApJ...862...98Z}, corresponding to the case of $L=40~\rm MeV$. and it gives $M_{\rm TOV} =2.08~M_\odot$ and $R= 11.77~\rm km$, $\Lambda=331$ for a $1.4\,M_\odot$ star. They are also demonstrated in Fig 5 to fulfill the black hole limit, the Buchdahl limit and the causality limit. We become aware that it agrees perfectly with some latest results within other theoretical framework, see, e.g., ~\cite{2018PhRvC..97b5806M,2018ApJ...860..139B,2019arXiv190104371M}.

\section{SUMMARY AND PERSPECTIVE}
In summary, in this contribution, we continued our recent work~\cite{2018ApJ...862...98Z} on NS EOS from the quark level in the light of GW170817. We confront our results with ab-initio calculations and find satisfying agreements. Important constraints on the parameter space of our model can be made especially from the chiral effective field study of neutron matter, namely the slope of the symmetry energy at saturation is found to be in the range of $30-60~\rm MeV$ within QMF. 

We also pay attention to the choice of crust-core matching and its possible influences on the GW tidal signals. We made plots for different $L$ values on the pressure as functions of the density. The over-simplified treatment of our work for the matching procedures, gluing different core EOSs to one crust model, prevents us discussing in a consistent way the effect of crust EOS. However, the present calculations seem to  clearly demonstrate that any claims regarding constraining the symmetry energy parameters with GW tidal signals should be considered with caution, although it may be safe to translate constraints on tidal deformability to constraints on the radius of merging stars. 

For future plans, along this line, we can make detailed studies for tidal deformability on the interplay of the saturation parameters with various possible strangeness phase transition~\cite{2018ApJ...860..139B,2018PhRvD..97h4038P,2018Most,2018arXiv180901116B,2018PhRvD..98f3020Z,2019PhRvD..99b3009C,2019xia} at higher densities (usually above $2\rho_0$). An extend $QMF18$ EOS with unified crust and core properties will be useful as well for supernova simulation or pulsar studies. The pulsar properties can be predicted~\citep{2016ApJS..223...16L} and updated studies can be done for short gamma-ray bursts~\citep{2016PhRvD..94h3010L,2017ApJ...844...41L}. 

Besides the proposed NS model of GW170817 in this contribution, the possibility of strange star merger for GW170817 has been tested and brings many new perspectives not only for this single event, but for the fundamental strong interaction, see, e.g., \cite{2018PhRvD..97h3015Z,2018RAA....18...24L,2018ApJ...852L..32D,2015ApJ...804...21G,2017IJMPS..4560042P}. Also, supramassive/hypermassive magnetars as the remnants of binary mergers if confirmed might put severe challenges to NS model since it requires the TOV mass of the underlying EOS should be no less than $\sim 2.3 M_{\odot}$, see discussions for example in \cite{2017ApJ...850L..19M,2018ApJ...852L..25R,2019MNRAS.483.1912P,2019ApJ...872..114S,2019arXiv190301466K}, although more decisive analysis from the electromagnetic counterparts are still necessary. Realistic nuclear EOSs seem hard to be beyond this value, see discussions in e.g., \cite{esym,eos,2018RPPh...81e6902B}. Our QMF model only gives a TOV mass around $2.1~M_{\odot}$, slightly lower than the APR one which is around $2.2~M_{\odot}$~\cite{apr}. It might be high time to resolve these tensions between microscopic many-body calculations of nuclear matter (with or without QCD transition~\cite{2015PhRvD..91d5003K,2018PhRvD..97b3018B}) and astrophysical NS merger observations.

\section{ACKNOWLEDGMENTS}
We thank Constança Provid{\^e}ncia and James Lattimer for enlightening discussions during the Xiamen-CUSTIPEN Workshop on the EOS of Dense Neutron-Rich Matter in the Era of Gravitational Wave Astronomy, Jan. 3-7, Xiamen, China. This work was supported in part by the National Natural Science Foundation of China (Grant No. 11873040).


\end{document}